\documentclass{PoS}

\usepackage{multirow}

\title{Potential Dark Matter Signals at Neutrino Telescopes}

\ShortTitle{Potential Dark Matter Signals at Neutrino Telescopes}

\author{\speaker{Marco Chianese}\thanks{I thank Stefano Morisi for fruitful discussion. Moreover, I acknowledge the JSPS KAKENHI Grant Number JP18H04340 that partially covered the costs to attend the ICRC 2019 conference.}\\
        Gravitation Astroparticle Physics Amsterdam (GRAPPA), Institute for Theoretical Physics Amsterdam and Delta Institute for Theoretical Physics, University of Amsterdam, Science Park 904, 1098 XH Amsterdam, The Netherlands\\
        E-mail: \email{m.chianese@uva.nl}}

\abstract{Recent analyses of the diffuse TeV-PeV neutrino flux highlight a tension between different IceCube data samples that strongly suggests a two-component scenario rather than a single steep power-law flux. Such a tension is further strengthened once the latest ANTARES data are also taken into account. Remarkably, both experiments show an excess in the same energy range (40-200 TeV), whose origin could intriguingly be related to dark matter. In this paper, I discuss the combined analysis of IceCube and ANTARES data, highlighting the presence of the low-energy excess. Moreover, I update the results of the angular analysis for potential dark matter signals, previously obtained with the 4-year High-Energy Starting Events data. In particular, I statistically compare the distribution of the arrival directions of 6-year IceCube events belonging to the low-energy excess with the angular distributions expected in case of different dark matter neutrino signals.}

\FullConference{36th International Cosmic Ray Conference -ICRC2019-\\
		July 24th - August 1st, 2019\\
		Madison, WI, U.S.A.}

\begin{document}

\section{Introduction}

The recent measurement of ultra-high energy (UHE) neutrinos at TeV--PeV energies by IceCube~\cite{Aartsen:2017mau} and ANTARES~\cite{Albert:2017bdv} telescopes has ushered us into a new era for astroparticle physics, providing an important diagnostic tool for physics and astrophysics~\cite{Ahlers:2018fkn}. So far, the IceCube detector has collected about 100 fully-contained neutrino events in the so-called High-Energy Starting Events (HESE) data set. In nine year of data-taking (2007--2015), the ANTARES telescope in the Northern hemisphere has additionally observed 33 events with energies above 20 TeV. Although these neutrinos cannot be explained in terms of the atmospheric neutrino background, their astrophysical origin is still not clear. High energetic neutrinos are expected to be produced in astrophysical cosmic-ray accelerators through the decay of charged pions, which are indeed originated in hadronic ($p$-$p$)~\cite{Loeb:2006tw,Murase:2013rfa} and photo-hadronic ($p$-$\gamma$)~\cite{Winter:2013cla,Murase:2015xka} interactions. In these processes, gamma-rays are produced as well through the decay of neutral pions. The production of both neutrinos and gamma-rays in astrophysical environments has been recently confirmed by the striking multi-messenger observation of the coincident gamma-ray and neutrino emission from the flaring blazar TXS 0506+056~\cite{IceCube:2018cha,IceCube:2018dnn}. However, it has been pointed out that blazar flares can only contribute up to about 10\% of the total observed neutrino flux~\cite{Murase:2018iyl}. Moreover, other searches for spatial and temporal correlations with gamma-rays~\cite{Adrian-Martinez:2015ver, Aartsen:2016oji, Aartsen:2018fpd,Aartsen:2018ywr} and for angular clustering~\cite{Aartsen:2014ivk,Ando:2017xcb,Mertsch:2016hcd,Dekker:2018cqu} have placed strong constraints on the contribution of several extragalactic astrophysical sources. Hence, the observed diffuse neutrino flux is expected to be given by a superposition of unresolved astrophysical sources. In this case, the neutrino flux is simply parameterized in terms of a diffuse power-law spectrum $E_\nu^{-\gamma_{\rm Astro}}$, where the key quantity is the spectral index $\gamma_{\rm Astro}$. 

The mystery of the origin of UHE neutrinos is further deepened by tensions among different IceCube data samples. Under the assumption of a single power-law flux, the 8-year through-going (TG) muon events provide a spectral index of $2.19\pm0.10$ as best-fit. On one hand, this is compatible with the theoretical expectation of the Fermi acceleration mechanism ($\gamma_{\rm Astro} = 2.0$) and with the measurements of the blazar TXS 0506+056, which suggest a hard power-law flux with a spectral index in the range $2.0\div2.3$~\cite{IceCube:2018dnn}. On the other hand, a spectral index of about 2.2 is in tension with the 6-year HESE data that instead require a steep power-law flux ($\gamma_{\rm Astro} = 2.92^{+0.29}_{-0.33}$). Such a discrepancy might suggest the presence of two components that dominate the diffuse neutrino flux at different energies and potentially have different angular properties~\cite{Aartsen:2017mau, Aartsen:2015knd}. Indeed, it is worth observing that the TG data sample collects muon neutrinos coming from the Northern hemisphere only and having energies larger that 200~TeV. Differently, the HESE events have a lower energy threshold of 20 TeV and cover the whole sky, so they are sensitive to the galactic centre of the Milky Way. Different analyses investigating the scenario with two astrophysical power-law components have pointed out that the data are compatible with the sum of a hard isotropic extragalactic neutrino flux and an additional softer one with a potential galactic origin (see, for example, Ref.s~\cite{Chen:2014gxa,Vincent:2016nut,Palladino:2016xsy,Palladino:2018evm}). Moreover, the two-component hypothesis is also supported by the first combined analysis of IceCube and ANTARES data~\cite{Chianese:2017jfa}. Remarkably, both IceCube and ANTARES telescopes have measured in the same energy range (about 40--200 TeV) a slight excess with respect to an astrophysical power-law flux deduced by TG data ($\gamma_{\rm Astro} \leq 2.2$), after the background subtraction~\cite{Chianese:2017jfa,Chianese:2016opp,Chianese:2016kpu,Chianese:2017nwe}.

In this context, heavy Dark Matter (DM) particles have been proposed as potential origin the diffuse UHE neutrino flux (see, for example, Ref.s~\cite{Chianese:2016opp,Chianese:2016kpu,Chianese:2017nwe, Feldstein:2013kka,Esmaili:2013gha,Esmaili:2014rma,Murase:2015gea,Boucenna:2015tra,DiBari:2016guw,Bhattacharya:2017jaw,Aartsen:2018mxl,Chianese:2018ijk,Bhattacharya:2019ucd}). In general, depending on the DM mass and the coupling between the dark sector and the ordinary matter, an observable flux of UHE neutrinos might be produced through the decays of DM particles. On the other hand, annihilating heavy dark matter does not provide in general a measurable neutrino flux given the unitarity limit on the cross-section. In this framework, the properties of a potential decaying DM neutrino signal can be studied by looking at the energy spectrum and the angular distribution of the observed neutrinos. Analyses mainly devoted to the neutrino energy spectrum are able to distinguish among different DM decay channels, which indeed provide different neutrino fluxes as a function of the neutrino energy. Moreover, different DM models are further discriminated once the constraints from gamma-rays data are considered in a multi-messenger approach. The current data (neutrinos and gamma-rays) show that hadronic final states are strongly disfavoured or excluded, while leptonic ones are slightly disfavoured or allowed~\cite{Cohen:2016uyg}. The most favourable case is a heavy DM candidate that is coupled only to neutrinos. For a particle physics perspective, models allocating neutrinophilic heavy DM candidates has some interesting features. For examples, Ref.~\cite{Chianese:2018ijk} points out that such models in general require the active neutrinos to be Dirac particles and a reheating temperature of the Universe of about 1~TeV in order to match the observed DM relic abundance.

Even though these energetic studies provide the general allowed features for DM particles, they are intrinsically not sufficient to claim the presence of a genuine DM signal in the neutrino data. This is mainly due to the uncertainty affecting the power-law neutrino flux produced by astrophysical sources. On the other hand, an astrophysical and a DM origin of UHE neutrinos can be discriminated by the fact that they have different angular distributions of the neutrino arrival directions. In particular, while the extragalactic astrophysical component is isotropic, a neutrino flux originated by DM particles is expected to have some correlation with the galactic centre of our galaxy. For this reason, analyzing the measured neutrino angular distribution is of paramount importance to potentially discover a dark matter signal.

\section{Tension with a single power-law neutrino flux}

\begin{figure}[t!]
    \centering
    \includegraphics[width=0.5\textwidth]{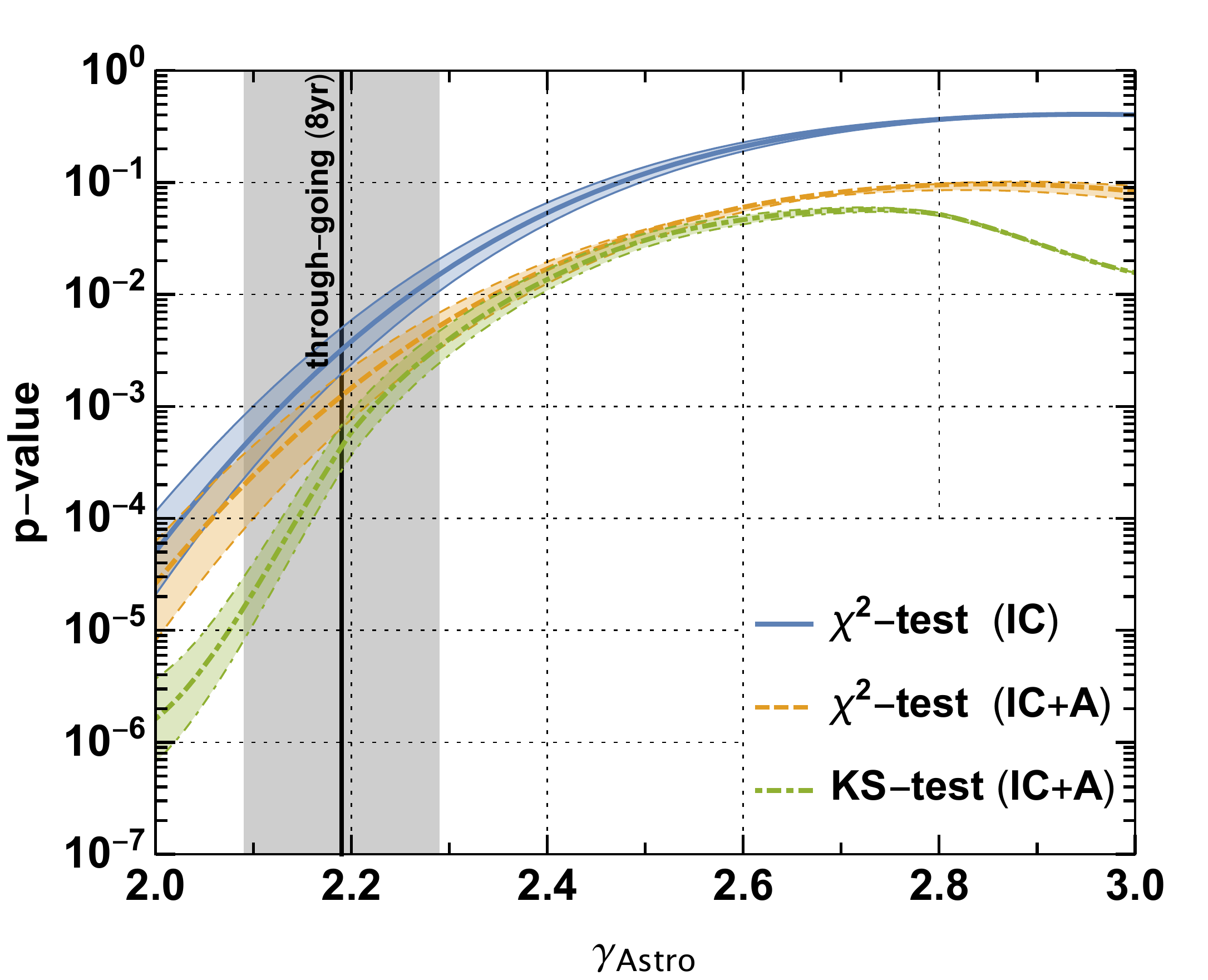}
    \includegraphics[width=0.4\textwidth]{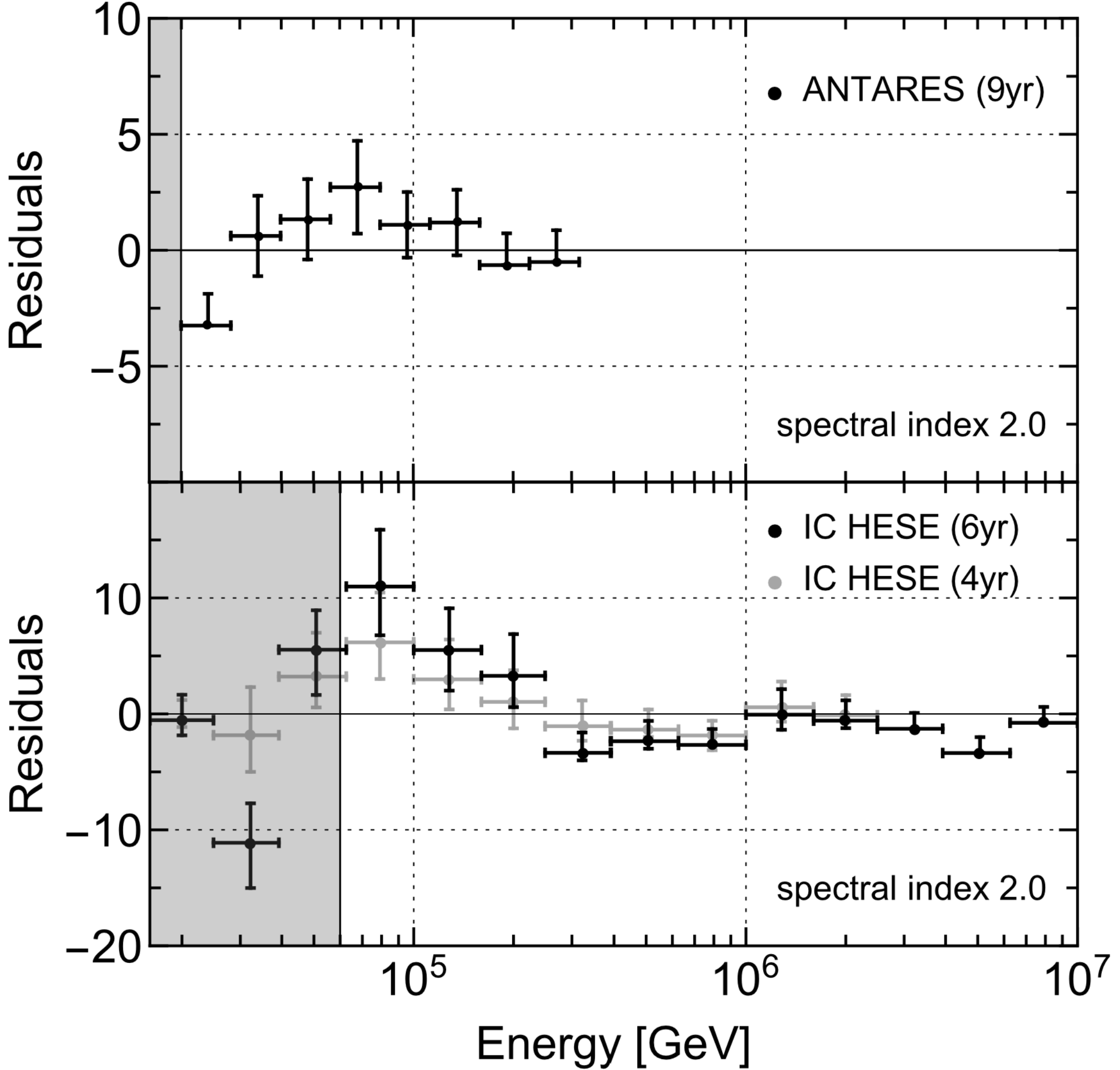}
    \caption{\textbf{Tension of neutrino data with a single power-law flux.} {\it Left}: The $p$-values as a function of the spectral index under the hypothesis of a single power-law flux. The different bands refer to different statistical tests ($\chi^2$ and Kolmogorov-Smirnov) and different data samples considered (IceCube only and IceCube+ANTARES). The vertical band represents the best-fit deduced by the IceCube 8-year through-going muon-neutrino data set. {\it Right}: ANTARES and IceCube residuals in the number of neutrino events with respect to the sum of the conventional neutrino background and an astrophysical power-law flux with spectral index of 2.0. The grey regions show the energy cuts considered in the fit of the power-law normalization.}
    \label{fig:1}
\end{figure}

As already discussed, the simplest explanation of the UHE neutrino flux is given by a single astrophysical power-law flux parameterized in terms of a normalization $\Phi_{\rm Astro}^0$ and a spectral index $\gamma_{\rm Astro}$. Such a flux is isotropic and has an equal flavor composition of neutrinos as expected for standard astrophysical sources. In order to quantify the tension of neutrino data with this assumption (considered as the null hypothesis), we perform a goodness-of-fit $\chi^2$ test and the non-parametric Kolmogorov-Smirnov one by considering the 6-year IceCube HESE data sample (IC) and the 9-year ANTARES one (A). In this analysis, we consider different values of the spectral index and, for each of them, the flux normalization (considered as free parameter) is determined by a maximum-likelihood procedure.

In the goodness-of-fit test, the two data samples are statistical combined by taking the product of the two likelihoods
\begin{equation}
    \chi^2 = -2 \ln \mathcal{L} \left(n^{\rm IC}, n^{\rm A} | \Phi_{\rm Astro}^0, \gamma_{\rm Astro} \right) = -2 \ln \mathcal{L}_{\rm IC}\cdot \mathcal{L}_{\rm A}\,,
    \label{eq:chi}
\end{equation}
where $n^{\rm IC}$ and $ n^{\rm A}$ are the number of neutrino events observed in IceCube and ANTARES, respectively. For each experiment, we consider binned multi-Poisson likelihoods. In each energy bin, we subtract the conventional atmospheric background while the prompt atmospheric background is considered to be negligible (see Ref.~\cite{Chianese:2017jfa} for details). For each value of the spectral index, the $p$-values are computed by considering that the test statistics~(\ref{eq:chi}) follows a $\chi^2$ distribution with $N-m$ d.o.f. with $m=1$ being the number of free parameters. Given the low energy cuts of the two experiments ($E_\nu\geq20$~TeV for ANTARES and $E_\nu\geq60$~TeV for IceCube), the total number of energy bins $N$ is 18.

The Kolmogorov-Smirnov (KS) test is based on the comparison between the empirical cumulative distribution function and the theoretical one expected in case of a single power-law flux (null hypothesis). In this case, the $p$-values are obtained by a bootstrap method for IceCube and ANTARES experiments, respectively, and later combined through the Fisher's method.

The results of such hypothesis tests are reported in the left panel of Fig.~\ref{fig:1}, where the bands take into account the uncertainty on the conventional atmospheric background. We find that combining the two data sets provides smaller $p$-values than the ones deduced by IceCube data only, independently of the spectral index. This implies that the range of spectral indexes disfavoured is enlarged once IceCube and ANTAREAS data are combined. Most importantly, we find that the power-law behaviour with $\gamma_{\rm Astro}=2.19\pm0.10$ deduced by the 8-year through-going muon-neutrinos has a $p$-value smaller that $2\times 10^{-2}$ for each hypothesis test considered. This means that the same interpretation for the 6-year HESE data only or the combined data set is statistically disfavoured. Finally, the benchmark model of Fermi acceleration mechanism ($\gamma_{\rm Astro}=2.0$) has a $p$-value in the range $2.6^{+3.6}_{-1.8}\times10^{-5}$ and $1.6^{+2.1}_{-1.0}\times10^{-6}$ for the $\chi^{2}$ and the KS statistical tests, respectively. A single astrophysical power-law component {\it \`a la} Fermi is therefore highly disfavoured by both data set. This is highlighted in the right panel of Fig.~\ref{fig:1} showing the residuals in the number of neutrino events with respect to the sum of the conventional atmospheric background and of an astrophysical power-law with $\gamma_{\rm Astro}=2.0$. Remarkably, the {\it low-energy} excess is present in both ANTARES and IceCube experiments in the energy range 40--200~TeV. The local maximum statistical significance of the excess is $\sim 2.6~\sigma$, larger than the one deduced by the previous 4-year HESE data set.

\section{Angular analysis for Dark Matter neutrino flux}

\begin{figure}[t!]
    \centering
    \includegraphics[width=0.75\textwidth]{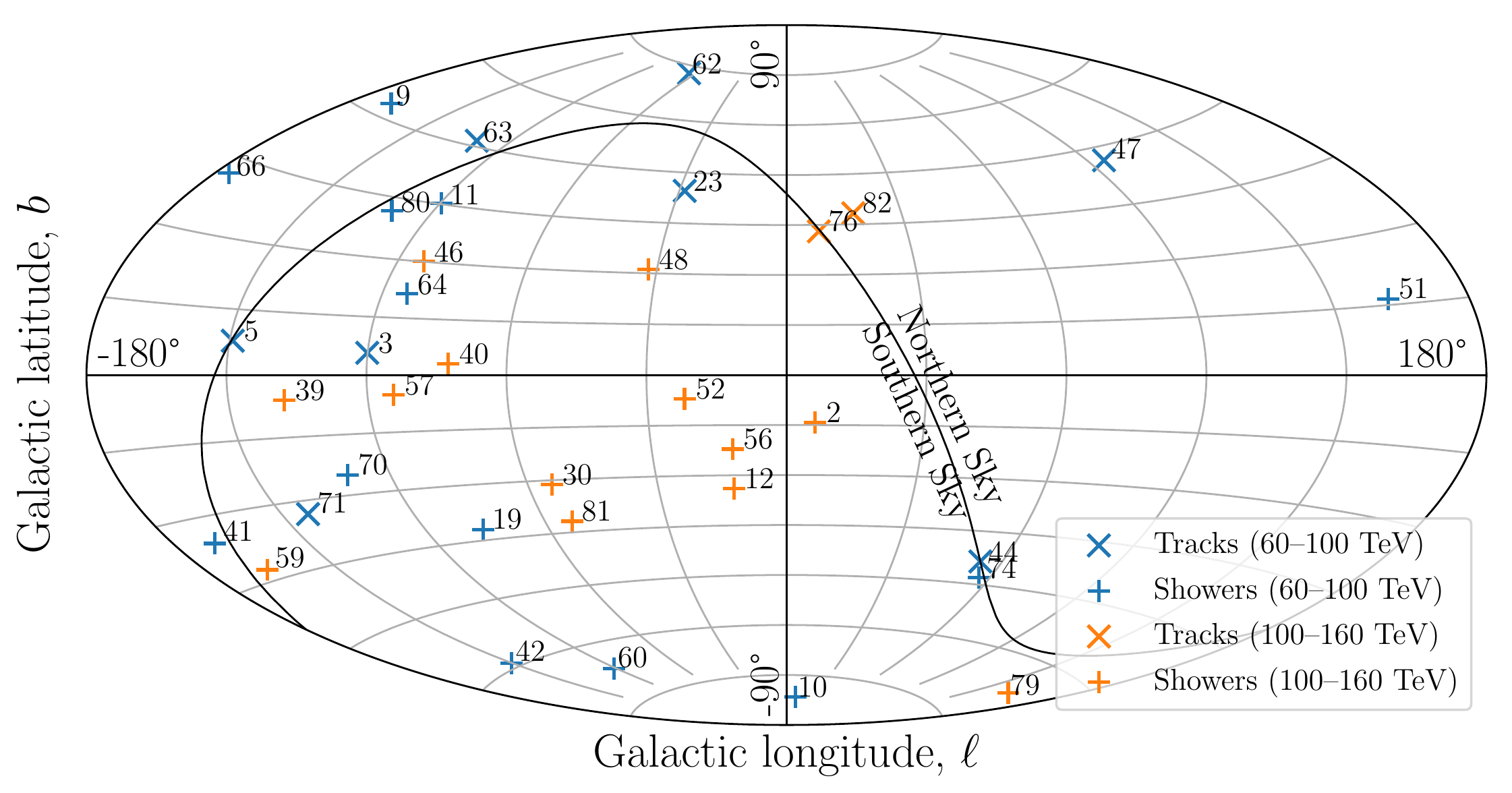}
    \caption{\textbf{Angular directions of the 6-year HESE neutrino events between 60 and 160~TeV.} The neutrino events are divided with respect to track ($\times$) and shower ($+$) topologies and to their energies, 60--100~TeV (blue) and 100--160~TeV (orange). Each neutrino event is identified by its IceCube number.}
    \label{fig:2}
\end{figure}

In order to infer the physical origin of the low-energy excess, we analyze the angular distribution of the observed neutrino events. In particular, the empirical angular distribution is compared with the expected angular distributions of a decaying or annihilating DM signal (see Ref.~\cite{Chianese:2016opp} for a similar study on galactic and extragalactic astrophysical sources). In both scenarios, the corresponding neutrino flux is expected to be peaked around the galactic centre where the DM density is in general higher. We update the results of the previous analysis focused on the 60--100~TeV energy bin of the 4-year HESE data by instead considering the latest public 6-year HESE ones. In this case, we combine two energy bins (60--100~TeV and 100--160~TeV) to almost cover the whole energy range of the low-energy excess, as shown in Fig.~\ref{fig:1}. The angular coordinates of the events considered in this analysis are displayed in Fig.~\ref{fig:2}. Due to the small statistics at our disposal, we assume that the excess is entirely explained by a DM component on top of the conventional neutrino background.

The expected angular distribution for DM neutrino signals has two contributions: a galactic one that depends on the DM density of our galaxy and an extragalactic one that is isotropic. Before considering the IceCube effective area, the non-normalized angular distributions for decaying and annihilating dark matter are respectively given by\footnote{For further details we refer the reader to Ref.~\cite{Chianese:2016opp}.}
\begin{eqnarray}
p^{\rm{dec}}(\cos\theta) & \propto &  \int_0^\infty\rho_h[r(s,\cos\theta)]{\rm d}s + \Omega_{\rm DM}\rho_c\,\beta_z \,, \\ 
p^{\rm{ann}} (\cos\theta) & \propto &  \int_0^\infty\rho^2_h[r(s,\cos\theta)]{\rm d}s + (\Omega_{\rm DM}\rho_c)^2\,\Delta^2_0 \,\beta_z\,.
\label{eq:dist}
\end{eqnarray}
The quantity $\rho_h$ is the DM galactic halo profile that depends on the line-of-sight $s$ and the angular variable $\cos\theta \equiv \cos b \cos l$, where $b$ and $\ell$ are the galactic coordinates. We consider the Navarro-Frenk-White (NFW) and the Isothermal (Isoth.) density profiles. The quantity $\Omega_{\rm DM} \rho_c$, instead, is the cosmological DM density. In case of annihilating scenario, the DM clumpiness is parameterized by the additional factor of $\Delta^2_0$. Furthermore, the dependence on the redshift $z$ is encoded in $\beta_z$. Such a quantity in general depends on the energy spectrum (DM channel) of neutrinos. However, as shown in Ref.~\cite{Chianese:2016opp}, under the assumption of a zero neutrino flux at high energies, there exists an upper bound for $\beta_z$. In particular, for the old and new analyses, $\beta_z\leq0.56/H_0$ and $\beta_z\leq1.07$ with $H_0$ being the today Hubble constant.
\begin{table}[t!]
\begin{center}
\begin{tabular}{|c|c||c|c||c|c|}
\hline
\multicolumn{2}{|c||}{\multirow{2}{*}{{\bf Scenario}}} & \multicolumn{2}{|c||}{\bf 4-year HESE (60-100~TeV)}& \multicolumn{2}{|c|}{\bf 6-year HESE (60-160~TeV)} \\
\cline{3-6}
\multicolumn{2}{|c||}{} & {{\bf KS}} & {{\bf AD}} & {{\bf KS}} & {{\bf AD}} \\ \hline\hline
\multirow{2}{*}{{{\small DM decay}}} & {\small NFW} & {\small0.06 - 0.16} & {\small0.03 - 0.14} & {\small0.07 - 0.58} & {\small0.13 - 0.60} \\ \cline{2-6}
& {\small Isoth.} & {\small0.08 - 0.22} & {\small 0.11 - 0.74} & {\small0.21 - 0.77} & {\small 0.05 - 0.19} \\ \hline \hline
{\small DM annih.} & {\small NFW} & {\footnotesize $\left(0.3 - 0.9\right)\times 10^{-4}$} & {\footnotesize $\left(0.3 - 3.8\right)\times 10^{-4}$} & {\footnotesize $\leq 3.6 \times 10^{-5}$} & {\footnotesize $\leq 3.0 \times 10^{-6}$} \\ \cline{2-6}
{\small $\Delta^2_0=10^4$} & {\small Isoth.} & {\footnotesize$\left(0.9 - 2.8\right)\times 10^{-3}$} & {\footnotesize$\left(1.0 - 5.0\right)\times 10^{-3}$} & {\footnotesize $\leq 5.7 \times 10^{-4}$} & {\footnotesize $\leq 9.9 \times 10^{-5}$} \\ \hline
{\small DM annih.} & {\small NFW} & {\small0.02 - 0.05} & {\small0.02 - 0.07} & {\small0.05 - 0.42} & {\small0.06 - 0.32} \\ \cline{2-6}
{\small $\Delta^2_0=10^6$} & {\small Isoth.} & {\small0.10 - 0.28} & {\small0.08 - 0.29} & {\small0.19 - 0.89} & {\small0.33 - 0.91} \\ \hline
\end{tabular}
\caption{{\bf Range of $p$-values for different Dark Matter scenarios and neutrino data sets.} The ranges correspond to a Monte Carlo on the angular neutrino errors, averaged over background configurations. The statistical tests are the Kolmogorov-Smirnov (KS) and the Anderson-Darling (AD) tests.}
\label{tab:tab1}
\end{center}
\end{table}

We perform two non-parametric one-dimensional statistical tests (Kolomogorov-Smirnov and Anderson-Darling) on IceCube data, under the assumption of different DM angular distributions as null hypothesis. The ranges of $p$-values obtained are reported in Tab.~\ref{tab:tab1}. They have been computed by considering the combinations of different choices of $n_{\rm bkg}$ background events among $n_{\rm tot}$ total number of events in each energy bin. For the 4-year HESE data set, in the energy bin 60-100~TeV we have $n_{\rm bkg}=5$ and $n_{\rm tot}=12$, so having 792 possible combinations. In the new analysis of 6-year HESE data, we instead consider two energy bins, 60-100~TeV and 100-160~TeV for which $n_{\rm tot}=21$ with $n_{\rm bkg}=8$ and $n_{\rm tot}=15$ with $n_{\rm bkg}=4$, respectively. Since we cannot scrutinize all the 203490 and 1365 possible background combinations in the two bins, we extract a sub-set of $10^5$ configurations. Moreover, in both analyses, in order to take into account the angular uncertainty, for each combination we consider 100 possible extractions of the observed events in their maximum error intervals. The $p$-values are computed by averaging over different combinations. Hence, the reported ranges cover all the background-averaged $p$-values obtained by the Monte Carlo concerning the experimental error.

As can be seen from the table, in general the updated analysis shows larger ranges and higher $p$-values than the previous results. The former is due to the fact that the number of combinations of signal events is increased and most of the events are showers so having large uncertainty on the angular reconstruction. The latter is instead due to the fact that, in the additional energy bin considered (100-160~TeV), there are few events close to the galactic centre (see Fig.~\ref{fig:2}) and therefore the DM scenarios are less constrained. However, the case of annihilating DM particles with a small clumpiness factor ($\Delta^2_0 = 10^4$) is still excluded with smaller $p$-values by the 6-year HESE data set.

\section{Conclusions}

In this paper, we have highlighted the tensions of the IceCube and ANTARES neutrino data with the simplest single astrophysical power-law model. The combined analysis of 6-year HESE IceCube and 9-year ANTARES data show that the astrophysical power-law behaviours deduced by the through-going muon neutrino data and the one expected by the standard Fermi acceleration mechanism are disfavoured (see Fig~\ref{fig:1}). This has suggested the presence of a second component in the neutrino flux dominating especially around 100~TeV where an excess is shown in both IceCube and ANTARES experiments (see Fig~\ref{fig:1}). We have tested the hypothesis that such a low-energy excess is due to a Dark Matter signal by analyzing the angular distribution of the arrival directions of neutrino events. As shown in Tab.~\ref{tab:tab1}, the current data (about 40 events between 60 and 160~TeV) are not powerful enough to constrain most of the Dark Matter scenarios considered (decaying and annihilating Dark Matter particles with different galactic halo density profiles and clumpiness factors). Only the annihilating DM scenario with a smaller clumpiness factor is excluded by data. As pointed out in Ref.~\cite{Chianese:2016opp}, a statistics of $\mathcal{O}(300)$ neutrino events is indeed required to exclude a Dark Matter origin of the low-energy excess.

\end{document}